\documentclass[12pt]{article}
\usepackage{graphicx}

\newcommand\pubnumber{NuPhys2017-Drakopoulou}
\newcommand\pubdate{\today}

\def\napoli{School of Physics and Astronomy\\
University of Edinburgh, EH9 3FD, Scotland, United Kingdom}

\def\Title#1{\begin{center} {\Large #1 } \end{center}}
\def\Author#1{\begin{center}{ \sc #1} \end{center}}
\def\Address#1{\begin{center}{ \it #1} \end{center}}

\newcommand\pubblock{\rightline{\begin{tabular}{l} \pubnumber\\
         \pubdate  \end{tabular}}}
\newenvironment{Abstract}{\begin{quotation}  }{\end{quotation}}
\newenvironment{Presented}{\begin{quotation} \begin{center} 
             PRESENTED AT\end{center}\bigskip 
      \begin{center}\begin{large}}{\end{large}\end{center} \end{quotation}}

\def\beq{\begin{equation}}
\def\eeq#1{\label{#1}\end{equation}}
\def\eeqn{\end{equation}}

\def\beqa{\begin{eqnarray}}
\def\eeqa#1{\label{#1}\end{eqnarray}}
\def\eeqan{\end{eqnarray}}

\let\bar=\overbar

\def\Dslash{\not{\hbox{\kern-4pt $D$}}}
\def\dslash{\not{\hbox{\kern-2pt $\del$}}}

\def\msb{{\bar{\ssstyle M \kern -1pt S}}}

\begin{document}
\begin{titlepage}
\pubblock

\vfill
\Title{ANNIE Phase II Reconstruction Techniques}
\vfill
\Author{ Evangelia Drakopoulou on behalf of the ANNIE Collaboration }
\Address{\napoli}
\vfill
\begin{Abstract}
The Accelerator Neutrino Neutron Interaction Experiment (ANNIE) \cite{ANNIE0} is a 26-ton Gd-doped water Cherenkov detector installed in the Booster Neutrino Beam at Fermilab. The experiment has two complimentary goals: (1) perform the first measurement of the neutron yield from $\nu_{\mu}$ interactions as a function of Q$^2$ in order to constrain neutrino-nucleus interaction theoretical models, and (2) demonstrate the power of new fast-timing, position-sensitive detectors by making the first deployment of Large Area Picosecond PhotoDetectors (LAPPDs) in a physics experi-ment. In Phase I, ANNIE successfully performed neutron background measurements. To realise the Phase II measurements the ANNIE collabo-ration has developed several reconstruction techniques using the arrival time and position of the Cherenkov photons in the detector photomultipliers (PMTs) and LAPPDs. A maximum-likelihood fit is used to reconstruct the neutrino interaction vertex and direction. Machine and Deep Learning techniques are used for the muon and neutrino energy reconstruction. We present the results of ANNIE reconstruction techniques and the improvement we can get in resolution with the use of LAPPDs \cite{ANNIE}. 
\end{Abstract}
\vfill
\begin{Presented}
NuPhys2017, Prospects in Neutrino Physics\\
Barbican Centre, London, UK,  December 20--22, 2017
\end{Presented}
\vfill
\end{titlepage}
\def\thefootnote{\fnsymbol{footnote}}
\setcounter{footnote}{0}

\section{The ANNIE Experiment}

The primary physics goal of ANNIE in Phase II is to study the multiplicity of final state neutrons from neutrino-nucleus interactions in water. These measurements will improve our understanding of the many-body dynamics of neutrino-nucleus intera-ctions and will allow to reduce the systematic uncertainties of the neutrino energy reconstruction in oscillation
experiments and the signal-background separation for neutrino experiments. Efficient detection of neutrons in ANNIE will be made possible by searching for a delayed signal from their capture on gadolinium (Gd) dissolved in water. Gd nuclei have high neutron capture cross-sections and produce 8 MeV gammas in several tens of microseconds after the initial event, which provide a detectable signal in water Cherenkov detectors. 

The ANNIE detector consists of a Gd-doped water detector deployed on the Booster Neutrino Beam at Fermilab \cite{BNB}. This beam is about 93\% pure $\nu_{\mu}$ (when running in neutrino mode) and has a spectrum that peaks at about 700 MeV. A Front Veto to reject entering backgrounds produced in the upstream rock and an external Muon Range Detector (MRD) downstream from the neutrino target are also utilised. The experiment is designed to proceed in two stages: a partially-instrumented test-beam run using only photomultiplier tubes (PMTs) (Phase I) for the purpose of measuring critical neutron backgrounds to the experiment \cite{ANNIE} and a physics run with a fully-instrumented detector (Phase II). 

Phase I is now completed and the analysis of the data collected ongoing.  ANNIE Phase II will be the first experiment to use fast-timing and position-precise LAPPDs \cite{LAPPD}, which are now being produced by Incom Inc., to perform these measurements [2]. To realise the physics goals for Phase II, the ANNIE collaboration has developed several track and kinematic reconstruction techniques.

\section{Track Reconstruction}
 
To reconstruct a beam neutrino interaction vertex, we employ an algorithm based on the techniques used in previous WCh detectors, such as Super-Kamiokande \cite{SK}. A track of a charged particle produced from a neutrino interaction can be characterized by six parameters: three spatial parameters specify the vertex position, one time parameter reflects when the interaction took place and two angular parameters specify the direction of the primary lepton track. A relativistic particle traveling in water will emit Cherenkov light, which is collected by the photodetectors mounted on the inner surface of the tank. The photon hit timing and the Cherenkov cone pattern are used to reconstruct the six parameters that are determined from a maximum-likelihood fit. These six parameters are varied in the fit to maximize the overall figure-of-merit (FOM), which is used to estimate the goodness of the fit. For the cone-edge fit, we build an analytical probability density function to describe the expected angular distribution of all digits. We then vary the track parameters and calculate an angular distribution from them, comparing with the PDF to give us the cone-edge component of the FOM. The best-fit vertex position and track parameters are those which maximize the overall FOM \cite{ANNIE}.

A sample of muons that are produced within the expected fiducial volume and stopped inside the MRD are selected for the event reconstruction. The vertex radial displacement ($\Delta$r) is defined as the distance between the reconstructed and the true vertex positions, and the track angular displacements ($\Delta\phi$) is defined as the angle between the reconstructed and the true track directions. Both the displacements are calculated on an event by event basis. We define the vertex (angular) resolution as the value of $\Delta$r ($\Delta\phi$) at the 68th percentile of all successfully reconstructed events from the sample. The vertex and angular resolutions are investigated for a PMT-only configuration including 128 8-inch PMTs (about 20\% coverage of the inner surface of the tank) and for a combined configuration including 128 PMTs and 5 LAPPDs on the downstream wall of the tank.

\begin{figure}[htb]
\centering
\includegraphics[height=1.8in]{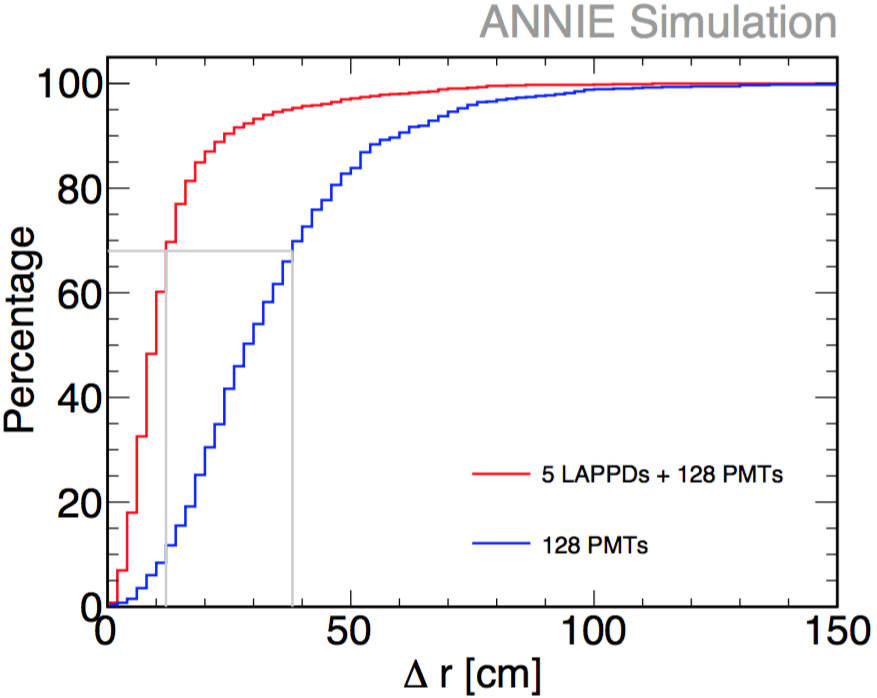}
\centering
\includegraphics[height=1.8in]{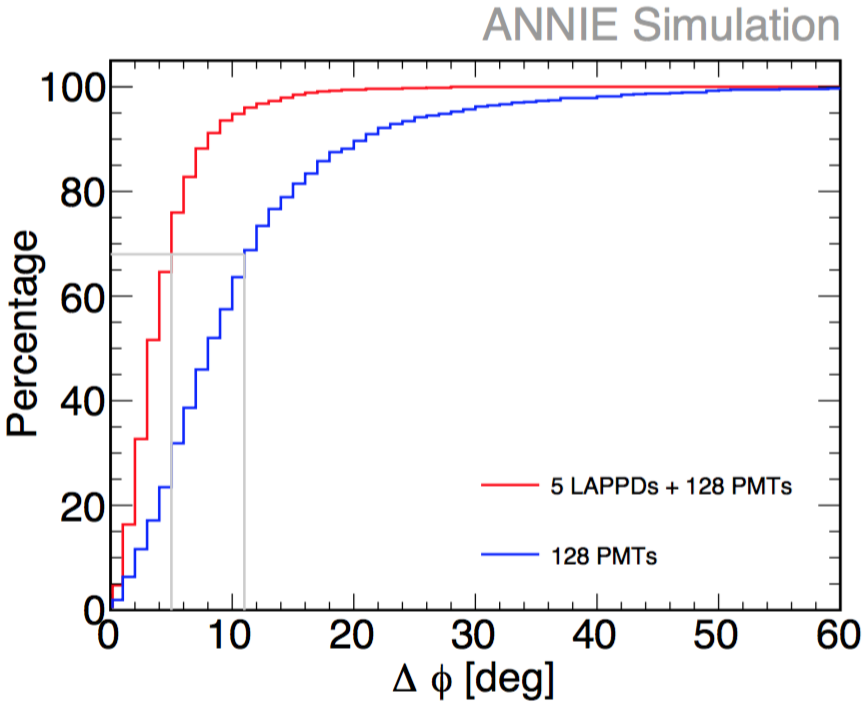}
\caption{Cumulative distributions of vertex (left) and direction (right) resolutions, for reconstructed events by the ANNIE detector with 128 PMTs only (blue) and 5 LAPPDs and 128 PMTs (red).}
\label{fig:R_phi}
\end{figure}

Figure \ref{fig:R_phi} shows the cumulative distributions, expressed as a percentage of successfully reconstructed events, for the vertex and direction reconstruction and compares the performance achieved in the two configurations. The vertex resolution from 20\% coverage with 128 conventional PMTs is about 38 cm. A configuration with 5 LAPPDs and 128 PMTs achieves a much improved resolution of 12 cm. Using the LAPPDs and PMTs combined configuration, the track angle can be reconstructed with a resolution of about 5 degrees, which is a factor of two improvement over the PMT-only configuration.

\section{Energy Reconstruction}
\begin{figure}[htb]
\centering
\includegraphics[height=1.7in]{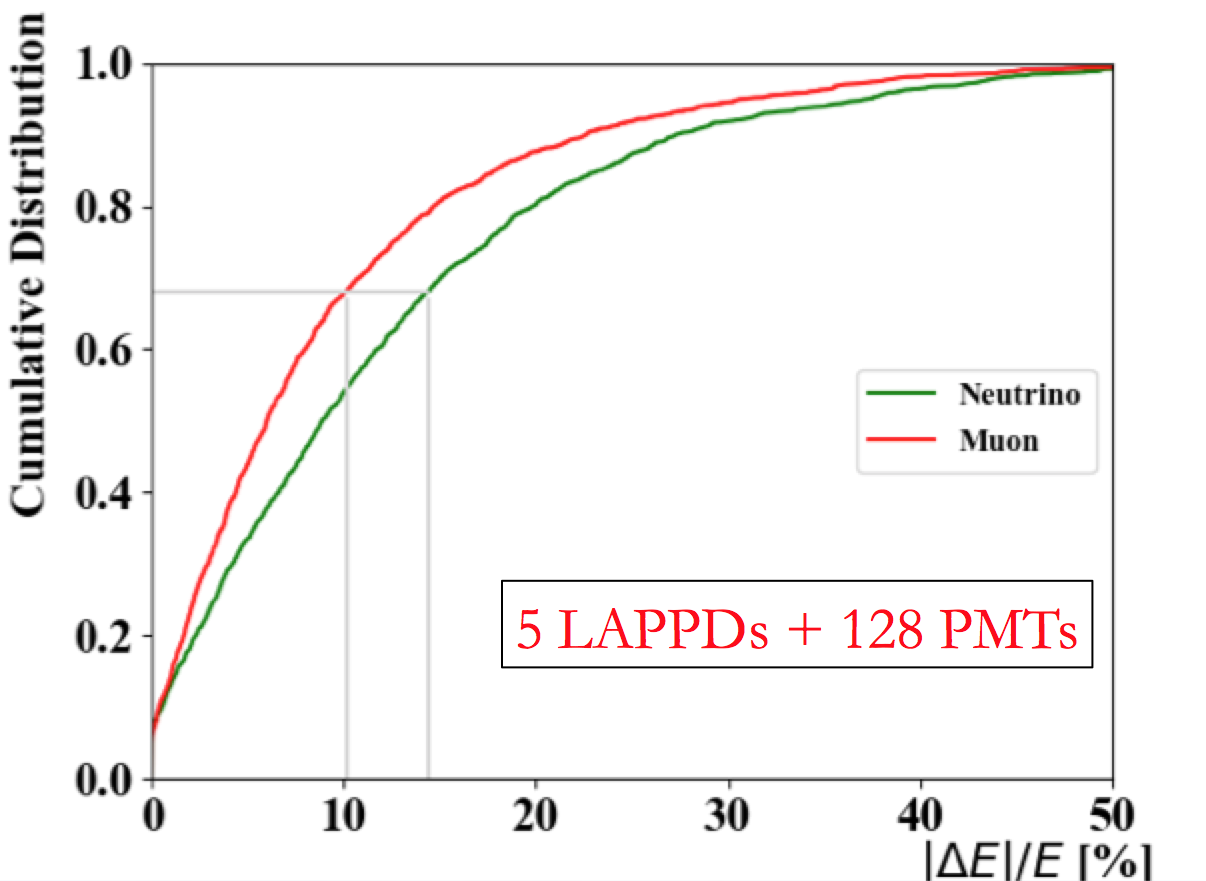}
\caption{The cumulative distribution of the muon (red) and neutrino (green) energy resolution for reconstructed events by the ANNIE detector with 5 LAPPDs and 128 PMTs.}
\label{fig:ener}
\end{figure}

To estimate the muon and neutrino energy quasi-elastic events stopped inside the MRD are selected. The track length in the MRD is calculated as a fit to the recorded hits. The track length in the water tank is reconstructed using a Deep Learning Neural Network algorithm (from Tensorflow 1.3.0). This is trained on multiple parameters including the photodetector hit times, the number of hits and the Cherenkov photons emission points based on the reconstructed interaction vertex and direction. The track length is then passed to a Boosted Decision Tree (BDT) together with the MRD track length, the reconstructed vertex position and angle, the distances of the reconstructed vertex from the detector walls and the number of hits in LAPPDs and PMTs for each event. The BDT (from Scikit-Learn 0.18.2), similar to that in \cite{EnerReco}, is used to reconstruct the muon and neutrino energy. Figure \ref{fig:ener} shows the muon and neutrino energy resolutions achieved for the ANNIE configuration with 5 LAPPDs and 128 PMTs. The energy resolution is defined as the percentage of $|\Delta E^{true} - \Delta E^{reco} |$/$\Delta E^{true}$. The muon (neutrino) energy resolution achieved at the 68th percentile of all reconstructed events from the sample is 10\% (14\%).

\section{Kinematic Reconstruction}

\begin{figure}[htb]
\centering
\centering
\includegraphics[height=1.8in]{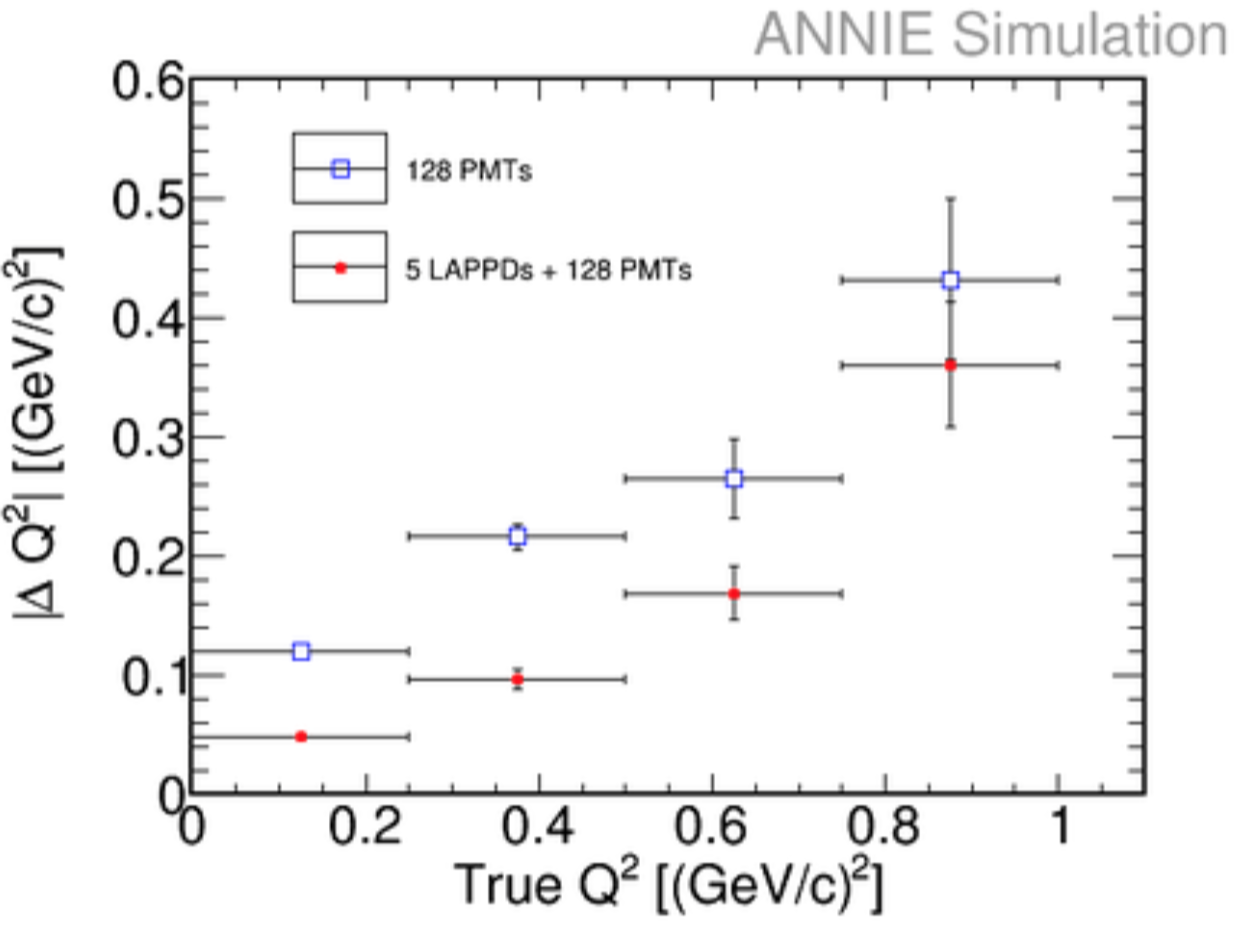}
\caption{The 1-$\sigma$ Q$^2$ resolution for four bins in true Q$^2$, reconstructed by the ANNIE detector with 128 PMTs only (blue) and 5 LAPPDs and 128 PMTs (red).}
\label{fig:Q}
\end{figure}

The reconstruction of event kinematics requires the estimation of the energy and direction of all particles in the final state. Present results have focused on quasi-elastic events, which are the primary interaction channel in ANNIE and are completely described by the energy of the incoming neutrino and the energy and momentum of the outgoing muon. Stopped muon events are selected for which the muon energy is measurable. The muon and neutrino energies from the BDT are used together with the reconstructed muon angle to calculate the momentum transferred, Q$^2$. Figure \ref{fig:Q} compares the accuracy of momentum transfer calculated for detector simulations using 128 PMTs alone, and with 128 PMTs and 5 LAPPDs. The addition of LAPPDs considerably improves the Q$^2$ resolution.

\section{Conclusions}
In Phase I, the ANNIE Collaboration performed measurements of the neutron backgrounds. In Phase II, the Collaboration plans to measure the neutron yield from $\nu_{\mu}$ interactions as a function of Q$^2$. The key technological component of Phase II, LAPPDs, are now being produced by Incom Inc. and the 
simulation and reconstruction tools for Phase II show good performance. The ANNIE Phase II data taking is foreseen in late 2018.


\end{document}